\documentclass[aps,prd,nopacs,nofootinbib,notitlepage]{revtex4-1}

\usepackage{amsfonts}
\usepackage{amsmath}
\usepackage{xcolor}
\usepackage{bm}
\usepackage{bbm}
\usepackage{graphicx}
\usepackage{fancybox}
\usepackage{comment}
\usepackage{multirow}
\usepackage{tipa}
\usepackage{longtable}
\definecolor{refs}{RGB}{245,156,74}
\usepackage[colorlinks=true,hyperfootnotes=true,citecolor=cyan]{hyperref}

\usepackage{enumitem}

\newcommand{\be}{\begin{equation}}
\newcommand{\ee}{\end{equation}}
\newcommand{\ba}{\begin{eqnarray}}
\newcommand{\ea}{\end{eqnarray}}
\newcommand{\bs}{\begin{subequations}}
\newcommand{\es}{\end{subequations}}

\newcommand{\e}{\mathrm{e}}
\newcommand{\ie}{\text{\textschwa}}

\newcommand{\diff}{\textrm{d}}

\newcommand{\lp}{\left(}
\newcommand{\rp}{\right)}

\begin{document}

\title{The canonical energy-momentum currents in cosmology}

\author{Tomi S. Koivisto}
\email{tomi.koivisto@ut.ee}
\address{Laboratory of Theoretical Physics, Institute of Physics, University of Tartu, W. Ostwaldi 1, 50411 Tartu, Estonia}
\address{National Institute of Chemical Physics and Biophysics, R\"avala pst. 10, 10143 Tallinn, Estonia}

\begin{abstract}

The parallel theory of relativity predicts conserved energy–momentum currents for an arbitrary metric, without invoking Killing symmetries. 
By treating the reference frame as an independent variational field and requiring it to carry no energy, the theory naturally unifies Einstein’s two formulations of gravity and yields uniquely defined covariant charges. 
In isotropic and homogeneous cosmology, the canonical time direction selected by the reference frame coincides with the Kodama vector, and the associated Noether energy reproduces the Misner–Sharp mass. 

\end{abstract}

\maketitle

\section{Introduction}

Physics is grounded in energy. The total energy of a system, its Hamiltonian, is the core quantity of classical mechanics and the single most important observable in the Dirac theory of quantum mechanics. 

However, in the presence of gravity the precise definition of energy has remained notoriously subtle.
The source of gravity in the standard formulation of General Relativity (GR) is the Hilbert 
energy-momentum tensor  $T^{\mu\nu}$, obtained from the variation of the source Lagrangian $L$ wrt the metric $g_{\mu\nu}$ as $T^{\mu\nu} = (2/\sqrt{-g})\delta (\sqrt{-g}L)/\delta g_{\mu\nu}$. 
This tensor does not by itself define a conserved current. Only in the presence of isometries, characterised by Killing vectors $\xi^\mu$, for which by definition\footnote{We shall denote geometric objects constructed from the metric with the ring. Thus, $\mathring{\nabla}_\mu$ is the Levi-Civita covariant derivative.} $\mathring{\nabla}_{(\mu}\xi_{\nu)} =0$, one then straightforwardly obtains properly conserved currents, as
\be
J^\mu_{\xi} = T^\mu{}_\nu\xi^\nu \quad \Rightarrow \quad \partial_\mu\lp \sqrt{-g}J_\xi^\mu\rp = 0\,.  
\ee
In the absence of a timelike Killing vector, there is no general procedure to find the conserved energy. In generic spacetimes, the dynamical geometry obscures the canonical pairing between time evolution and energy that underpins Hamiltonian dynamics.

Spherically symmetric spacetimes provide a notable exception.
In this case, there exists a widely accepted notion of energy, the Misner–Sharp mass \cite{Misner:1964je}. 
Under spherical symmetry, one can define the Kodama vector $K^\mu$ \cite{Kodama:1979vn} which, despite $\mathring{\nabla}_{(\mu}K_{\nu)}\neq 0$, introduces the identically conserved current $J_K^\mu = -\mathring{G}^\mu{}_\nu K^\nu$ that can be associated with a Komar superpotential \cite{Komar:1958wp} whose surface integral coincides with the Misner-Sharp mass  \cite{Hayward:1994bu}. More general quasilocal charge constructions \cite{Maeda:2007uu} are possible within the Iyer-Wald formalism \cite{Iyer:1994ys} via Hamiltonian corrections \cite{Regge:1974zd} to the Komar superpotential.
Beyond spherical symmetry, however, no comparably well-established local timelike vectors, conserved currents, or energy charges are known in GR. 

In the framework dubbed General Parallel Relativity (G$_\parallel$R) \cite{Gomes:2023hyk}, this obstruction is bypassed in a constructive manner. 
The timelike vector $\ie_0{}^\mu$ is canonically determined for an arbitrary metric, 
such that the current $J_0^\mu = -T^\mu{}_\nu\ie_0{}^\nu$ is conserved on shell and thus defines a unique, physically motivated notion of energy. 
In fact, G$_\parallel$R determines the quartet $\{\ie_a{}^\mu\}_{a=0,1,2,3}$ of such vectors for any given metric, corresponding to the general-relativistic energy and momenta. These energy and momenta are obtained from the theory's first-order action functional as the $\ie_a$-generated Noether charges \cite{BeltranJimenez:2021kpj}. The `Kodama miracle' \cite{Abreu:2010ru} $\mathring{\nabla}_\mu J_K^\mu =0$, which occurs in GR in the absence of an underlying symmetry principle, is understood as a special case for $J_K = J_0$ amongst the genuine Noether currents $J_a$ in G$_\parallel$R. 

The G$_\parallel$R theory has emerged recently \cite{Gomes:2023hyk}, building on developments scattered across the literature since 
\cite{BeltranJimenez:2017tkd}.  
This paper aims to provide a compact and accessible presentation of the theory, illustrated with a concrete example drawn from cosmology.  
We begin with a precise definition of charges in Sec.\ref{noether}.  
In Sec.\ref{GPR} it is shown how G$_\parallel$R unifies and completes Einstein's two formulations of GR. 
Sec.\ref{cosmo} applies the generic realisations of cosmological frames found by Beltr{\'a}n \cite{Gomes:2023hyk}, elaborates the interpretation of the canonical frame \cite{BeltranJimenez:2019bnx}, and reports the conserved currents and their integrated charges in cosmology. We conclude with Sec.\ref{conclu}.    

\section{Charges}
\label{noether}

We begin with a general discussion of charges associated with conserved currents. The conservation of a densitised current \(\sqrt{-g}J^{\mu}\) is equivalently expressed in the covariant form 
\be \label{cons}
\partial_{\mu}\left( \sqrt{-g}\, J^{\mu} \right) = 0
\quad \Leftrightarrow \quad
\mathring{\nabla}_{\mu} J^{\mu} = 0 \,.
\ee
We consider charges in a spatial three-volume $V$. 
This requires a \(1+3\) decomposition. For concreteness, let us adopt the standard ADM decomposition \cite{Regge:1974zd} in which the region $V$ is described by the spatial metric $h_{ij}$ induced from $g_{\mu\nu}$, together with the lapse $N$ and the shift $N^i$, so that the line element reads
$\diff s^2 = -N^2\diff t^2 + h_{ij}(\diff x^i +N^i\diff t)(\diff x^j + N^j\diff t)$. The naive charge would then be 
\bs
\label{NVcharge}
\be \label{Ncharge}
\tilde{Q}[V] = \int_{V}\diff^3 x \sqrt{h} J^t\,.
\ee
It is well-known that the conservation of the charges depends on the boundary conditions, in particular, whether there is a flux through the boundary $\partial V$. The subtlety here is that this is not implied by (\ref{cons}) for the naively defined charge (\ref{Ncharge}), except in special cases\footnote{In Hamiltonian formulation of GR, the lapse 
is an essential part of the generators and of the associated charges \cite{Regge:1974zd}. In contrast, conserved currents and observables are often considered implicitly in gauges with $N=1$.} 
where $\partial_\mu N=0$. One should consider instead
\be \label{Vcharge}
{Q}[V] = \int_{V}\diff^3 x N\sqrt{h} J^t \,, 
\ee
\es
for which one readily obtains the flux law
\be
\frac{\diff Q[V]}{\diff t} =  -\oint_{\partial V}\diff^2 x  N\sqrt{\gamma} J^{i} n_{i} \,,  
\ee
where $n_i$ is the outward-pointing unit normal, $h^{ij}n_i n_j=1$, to the boundary $\partial V$, and $\gamma$ is the determinant of
the two-metric induced from \(h_{\mu\nu}\) on \(\partial V\). As expected, \(Q[V]\) depends not only on the spatial
region but also on the choice of time coordinate via the lapse \(N\). 

Another subtlety is that physical observables are operationally defined by surface
charges rather than volume integrals such as (\ref{NVcharge}). Empirically,
measurements probe the fields excited by sources, since the presence of a
source can only be inferred through its interactions. 
Using the Gauss-Amp\`ere law $\mathring{\nabla}_\mu F^{\mu\nu}=J^\nu$ and
Stokes' theorem, the volume charge (\ref{Vcharge}) can be expressed as the
surface integral
\begin{equation}\label{Scharge}
q[V] = \oint_{\partial V} \mathrm{d}^{2}x N \sqrt{\gamma} \, F^{it} n_i \, .
\end{equation}
Once again, the naive definition (\ref{Ncharge}) fails to reproduce this
relation. The new point here is that the `local charge' (\ref{Vcharge}) is
equivalent to the `quasilocal charge' (\ref{Scharge}) only when the 
topology permits a direct application of Stokes'
theorem. 
In many physically relevant cases this is not so. For example, for the Dirac magnetic monopole one has 
$q \neq Q=0$, clearly demonstrating that the surface charge (\ref{Scharge}) provides the more fundamental definition of an observable.

While illustrated here using the Gauss-Amp\`ere law of
electromagnetism, the underlying structure is universal: a source is the
divergence of a field. By the Poincar\'e lemma, any conserved current $J^\mu$ satisfying (\ref{cons}) can be 
locally written as a divergence, $J^\mu = \mathring{\nabla}_\nu j^{\mu\nu}$, with $j^{\mu\nu}$ antisymmetric.
The tensor $j^{\mu\nu}$ is determined only up to the addition of a further
divergence, but such ambiguities integrate to zero on a closed surface and
therefore do not affect the quasilocal charge (\ref{Scharge}).

\section{Gravity}
\label{GPR}

This section demonstrates how a canonical notion of energy-momentum arises in gravity. This is achieved by subjecting the reference frame to a 
boundary-restricted variational principle and by requiring that it carries no intrinsic energy. These two principles -- aligned with the principles of equivalence and relativity, respectively -- lead to an essentially unique completion of GR. 

To begin, we distinguish coordinate symmetry (Diff) from a frame symmetry (GL). 
Diffeomorphisms refer to spacetime localisation, but operational physical measurements also involve local transformations of reference frames.
Whereas a position in spacetime is labelled by 4 coordinates, a reference spacetime geometry requires 4$^2$ = 16 components. We encode these in the quartet of vectors $\ie_a{}^\mu$, where Greek indices transform Diff-covariantly and Latin indices GL-covariantly. The dual coframe $\e^a{}_\mu$ satisfies $\e^a{}_\mu\ie_a{}^\nu=\delta^\nu_\mu$, $\e^a{}_\mu\ie_b{}^\mu=\delta^a_b$. We shall incorporate the frame into the theory as an independent variational field, besides the metric tensor field. Whereas the Diff-invariance of the action formulation is taken for granted, we shall not require exact GL-invariance: observables are frame-dependent in the operational sense that measurements are performed in a chosen reference frame, even though the underlying field equations remain Diff-covariant. In particular, the action functional underlies the partition function, and the frame-dependence of its boundaries is therefore a physical feature.
Thus, we shall consider an action principle which is a Diff-scalar but GL-invariant only up to a total derivative. 

The two fields induce their respective covariant derivatives: the metric-compatible $\mathring{\nabla}_\alpha g_{\mu\nu}=0$ and the frame-compatible $\nabla_\alpha\e^a{}_\nu=0$. This reflects the fact that gravitational free fall and the inertial motion defined by the physical reference frame need not coincide a priori. 
The difference of the associated connection coefficients is a tensor. From quadratic contractions of this  `distortion tensor' we can construct a scalar action functional $I[\ie,g]$ \cite{BeltranJimenez:2019odq}:
\be \label{GPRaction}
I[\ie,g] = \kappa\int\diff^4 x\sqrt{-g} X^\alpha{}_{[\beta}{}^\beta X^\gamma{}_{\gamma]\alpha}\,, \quad \text{where} \quad
X^\alpha{}_{\mu\nu} \equiv  \ie_a{}^\alpha\partial_\mu\e^a{}_\nu - \mathring{\Gamma}^\alpha{}_{\mu\nu}\,.  
\ee 
The coupling constant $\kappa=1/8\pi G$ with $G$ the Newton's constant. 
The tensor $X^\alpha{}_{\mu\nu}$ encodes the deviation between the gravitational free-fall motion determined by the metric and the acceleration of the reference frame, and thereby encodes the inertial content relevant for the definition of energy. 
If we enforce $\e^a{}_\mu=\delta^a_\mu$ (the so-called coincident gauge \cite{BeltranJimenez:2022azb}), the action is a functional of the metric alone, $I=I[g]$. This reproduces precisely Einstein's original action (i.e. the `Hamiltonian function') for GR\footnote{Covariant extensions of Einstein's original formulation in terms of a reference connection have been considered previously in e.g. \cite{Tomboulis:2017fim}.} \cite{Einstein1916}. Because of its explicit dependence on the non-tensorial $\mathring{\Gamma}^\alpha{}_{\mu\nu}$, $I[g]$ is not a scalar. The resulting description of gravity, modelled in analogy with electromagnetism, therefore relied on pseudotensors for the gravitational energy-momentum. 
While extremely successful for the dynamics of $g_{\mu\nu}$, it failed to provide a satisfactory covariant description of energy. 

An alternative formulation is obtained by taking the frame field as the only independent variational degree of freedom.
The metric $g_{\mu\nu}$ is then eliminated by introducing a constant Minkowski metric $\eta_{ab}$ and enforcing $g_{\mu\nu}=\eta_{ab}\e^a{}_\mu\e^b{}_\nu$ (the so-called metrical frame). 
The resulting action $I[\ie]$ reproduces precisely Einstein's second formulation of GR \cite{Einstein1928}. 
Although originally motivated by an attempt to unify gravity and electromagnetism, the frame formulation has since resurfaced in discussions of the localisation of gravitational energy \cite{Moller1958}. 

In this way, the action (\ref{GPRaction}) of G$_\parallel$R brings together Einstein's two formulations of GR\footnote{The promotion of $I[\ie]$ and $I[g]$ into Lorentz- and Diff-invariant theories, respectively, together with the Hilbert action $I_H=-(\kappa/2)\int\diff^4 x\sqrt{-g}\mathring{R}$, gives rise to the `Geometrical Trinity of Gravity' \cite{BeltranJimenez:2019esp}.}.  By construction, the generalised action $I[\ie,g]$ retains the relation $\kappa\mathring{G}_{\mu\nu}= T_{\mu\nu}$ that links matter to curvature. Conceptually, this celebrated relation beautifully realises what may be termed the Riemann-Clifford principle, but it 
does not fully capture the relativity principle dating back to Galilei. The latter concerns the physical role of reference frames.  
It is precisely the additional presence of the frame field that appears essential for fulfilling the original physical aims of GR and, further, for formulating a general-relativistic quantum theory. 

The frame-dependence of the action has been interpreted as a manifestation of the equivalence principle \cite{Koivisto:2018aip}. {\it Requiring that the frame does not contribute to dynamics determines the dynamical laws}, in the sense that the above $I[\ie,g]$ is the unique quadratic action whose frame variation only occurs at the boundary \cite{BeltranJimenez:2019odq}. 
In other words, by postulating trivial equations of motion for the field $\ie_a{}^\mu$ we may deduce the equations of motion for the field $g_{\mu\nu}$. 
Nevertheless, the field $\ie_a{}^\mu$ has, in general, a non-vanishing local energy current. 
The principle of relativity distinguishes a canonical class of reference frames characterised by vanishing frame energy \cite{Koivisto:2022nar}.
This fixes the configuration of the field $\ie_a{}^\mu$. {\it Requiring that the frame does not contribute to observations determines the observables}. 
In other words, by postulating trivial energy-momentum for the field $\ie_a{}^\mu$ we may deduce the reference frame wrt which to measure the physical energy-momenta of the other fields in the theory.  

Finally, we state explicitly the two defining equations of the theory: 
\be \label{GPReq}
\ie_a{}^\nu \kappa\mathring{G}^\mu{}_\nu = \mathring{\nabla}_\nu F_a{}^{\nu\mu} = J_a{}^\mu\,, \quad \text{where} \quad
F_a{}^{\mu\nu} = -\kappa\ie_a{}^\alpha\lp X^{[\mu}{}_\alpha{}^{\nu]} + 2\delta^{[\mu}_\alpha X^{\nu]}{}\rp\,, \quad X^\mu \equiv X^{[\mu}{}_\alpha{}^{\alpha]}\,. 
\ee 
The first equation is equivalent to the vanishing of the local canonical energy-momentum current associated with the frame field \cite{Gomes:2023hyk}. 
Hence, the first equation is a geometrical expression of the principle of relativity: it selects the canonical, energy-free frame for a given metric. The second equation expresses the dynamical law in its canonical form  \cite{Koivisto:2019ggr}: the divergence of the gravitational field $F_a{}^{\mu\nu}$ is the source energy-momentum current $J_a{}^\mu = T^\mu{}_\nu\ie_a{}^\nu$. The energy-momentum charges are thus, according to our general discussion in Sec.\ref{noether}, 
\be \label{GPRcharge}
q_a = \oint_{\partial V}\diff^2 x N\sqrt{\gamma} F_a{}^{t i }n_i\,.
\ee
The same expression (\ref{GPRcharge}) is obtained unambiguously from the action (\ref{GPRaction}) through the Noether procedure \cite{BeltranJimenez:2021kpj,Gomes:2022vrc,Koivisto:2022oyt}. We now proceed to test this construction explicitly in the framework of isotropic and homogeneous geometry.

\section{Cosmology}
\label{cosmo}

This section provides an explicit realisation of the canonical frame in a concrete spacetime, illustrating how the abstract construction of Secs.\ref{noether}-\ref{GPR} operates in practice and leads to physically meaningful charges. An isotropic and homogeneous spacetime lacks a timelike Killing vector, and can therefore accommodate some nontrivial features of the G$_\parallel$R framework.  

We adopt a 1+3 decomposition of the metric, $g_{\mu\nu} = h_{\mu\nu} - u_\mu u_\nu$, in terms of a timelike 4-velocity $u_\mu u^\mu = -1$ and the spatial projector $h_{\mu\nu}$. The most general isotropic and homogeneous metric with a spatial curvature $k$, first considered by Friedmann and by Lema\^itre, is then described in the comoving Cartesian coordinates by 
\be \label{frw}
u = N^{-1}\partial_t\,, \quad h^{ij} = a^{-2}\lp \delta^{ij} - k{x^i x^j}\rp\,.   
\ee
There are two time-dependent functions, the lapse $N$ and the scale factor $a$. 

The generic isotropic and homogeneous frames $\ie_a{}^\mu$, constructed by Beltr{\'a}n \cite{Gomes:2023hyk}, likewise involve two time-dependent functions, $\sigma$ and $\mu$, together with a length scale $\ell$. A priori, all these are independent of the two metric functions and the length scale $1/\sqrt{k}$. This highlights that the reference frame carries information beyond the metric geometry. 
As discussed in Sec.\ref{GPR}, the metric and the frame represent mathematically distinct structures, and their physical relation is fixed only after solving the equations of G$_\parallel$R. 

Since the frames are subject to both Diff and GL transformations, the symmetry can be realised by a mixed action of the two, see e.g. \cite{Nicolis:2015sra}. 
Isotropy and homogeneity of the frames admit two such inequivalent realisations, distinguished by how the Diff and GL transformations are intertwined.
\begin{itemize}
\item In the `regular' realisation, the coframe is given as
\bs 
\label{regular}
\ba 
\e^0 & = & \mu\chi\diff t - \sigma\chi^{-1}k\ell x_i\diff x^i\,, \\ 
\e^i & = & \mu\ell^{-1}x^i\diff t + \sigma\diff x^i\,,
\ea
\es
with $\chi = \sqrt{1-kr^2}$ where $r^2 = \delta_{ij}x^ix^j$. The regular case trifurcates into different realisations of flat cosmology in the limit $k \rightarrow 0$. This limit
can be seen as a group contraction that can retain $\ell$ finite, or let $\ell \rightarrow \infty$ with the energy scale $k\ell$ then either remaining finite or shrinking to zero. A certain duality relates the first and third cases, but all three are inequivalent.   
\item In the `exceptional'  realisation, the coframe is of the form 
\bs
\label{exceptional}
\ba
\e^0 & = & \mu\diff t\,, \\
\e^i & = & \sigma\lp \chi \diff x^i + \ell^{-1}\epsilon^i{}_{jk}x^k\diff x^j\rp\,. 
\ea
\es
In this case, only a single SO(3) $\in$ GL(4) is used to realise both homogeneity and isotropy, in conjunction with the respective Diffs. As the symmetry thus recovered is SO(3)$\times$SO(3) $\simeq$ SO(4), this realisation only applies to closed $k>0$ cosmologies. 
\end{itemize}
This unified picture of cosmological frames is likely to lead to further applications\footnote{The classification of frames is reflected in the five branches of cosmological solutions for the teleparallel connection $\Gamma^\alpha{}_{\mu\nu} =  \ie_a{}^\alpha\partial_\mu\e^a{}_\nu$ presented in \cite{Heisenberg:2022mbo}. Understanding \cite{Gomes:2023hyk} the three `regular' realisations of flat cosmology in symmetric teleparallel gravity models \cite{Hohmann:2021ast} was instrumental to uncover their degrees of freedom \cite{Gomes:2023tur}. The `exceptional' realisation corresponds to the special case of Witten’s Ansatz \cite{Witten:1976ck} as adapted to cosmology in \cite{Galtsov:1991un}, and applied also more recently in \cite{Murata:2021vnb}.}. 

However, here we only work with the canonical frame \cite{BeltranJimenez:2019bnx}, i.e. the solution to the G$_\parallel$R equations (\ref{GPReq}). Using the Ans{\"a}tze (\ref{regular}) and (\ref{exceptional}) together with the metric (\ref{frw}) for the first equation (\ref{GPReq}), it yields the solution exclusively for the `regular' case (\ref{regular}). The two unknown frame functions are then determined in terms of the two metric functions as  $\mu=\mu_0 N$, $\sigma=\mu_0/(\ell H)$, where $\mu_0$ is a dimensionless constant
and $H$ is the expansion rate $H \equiv u^\mu\partial_\mu\log{a} \equiv \log{a}'$ defined invariantly wrt time reparameterisations. 
The solution for flat cosmology fixes the consistent contraction limit such that the independent length scale $\ell$ is retained as $k \rightarrow 0$. 
Explicitly, we have in terms of the coframe and the frame, respectively, 
\bs
\label{thesolution}
\ba
\e^0  & = & N\chi\diff t - \frac{k}{\chi H}x_i\diff x^i\,, \quad
\e^i  =  \ell^{-1}\lp N x^i\diff t + \frac{1}{H}\diff x^i\rp\,, \\
\ie_0 & = & \frac{\chi}{N}\partial_t - \chi H x^i\partial_i\,, \quad
\ie_i = \ell\lp \frac{k}{N}x_i\partial_t + H\partial_i - H kx_ix^j\partial_j\rp \,,  \label{canonicalframe}
\ea
\es
with $\mu_0 = 1$ for simplicity.  The four vectors (\ref{canonicalframe}) set up the
canonical frame singled out by G$_\parallel$R.  The timelike vector
$\ie_0$ is tangent to the preferred cosmological flow. For $k=0$ it reduces to the
comoving four–velocity $\ie_0 \rightarrow u$, and for a general $k$ it coincides, $\ie_0=K$, 
with the Kodama vector $K$ available in the context of spherically symmetric metrics \cite{Kodama:1979vn}.  
It is worth stressing that the Kodama vector is not assumed here but emerges uniquely as the canonical time direction 
selected by the theory in accordance with the general construction in Sec.\ref{GPR}. 

The three vectors $\ie_i$ form a distinguished
spatial triad. In the flat case their time components vanish and the spatial parts reduce to
the Hubble-rescaled Cartesian translations $\ie_i \rightarrow \ell H\partial_i$. In general, 
they play the role of transvection generators which, together with the rotation 
generators $L^i=\epsilon^{ijk}x_j\partial_k$, satisfy the algebra
\be
[L_i,L_j] = \epsilon_{ijk} L^k\,,\quad
[L_i,\ie_j] = -\,\epsilon_{ijk} \ie^k\,, \quad
[\ie_i,\ie_j] = \ell^2 k\lp H^2  + H' \rp\epsilon_{ijk} L^k\,.
\ee
This algebra makes explicit how cosmological expansion deforms spatial translations into time-dependent transvections in the presence of curvature. 
The part of the frame algebra that mixes the transvections into rotations is controlled by the (squared) ratio of the two length scales and by the acceleration of the scale factor (given by the active gravitational mass density). The time evolution of these generators is reflected in the commutators
\be
[L_i,\ie_0] = 0\,, \quad [\ie_0,\ie_i] = - H\ie_i\,. \label{0icommutator}
\ee
The first relation confirms that the canonical time flow preserves isotropy. The second relation shows that the transvection generators are `redshifted' by the Hubble expansion, encoding the extrinsic curvature $\mathring{K}_{\mu\nu} = H h_{\mu\nu}$ via a generalisation of the Weingarten map $[u,X]^\mu=-\mathring{K}^\mu{}_{\nu}X^\nu$ for a spatial vector $X^\mu$. Since the transvections $\ie_i$ are not strictly tangent to the constant-time hypersurfaces for $k\neq0$, the commutator with the canonical time flow $\ie_0 \neq u$ realises a \emph{generalised} Weingarten (shape) operator, extending the classical one defined only on constant-time hypersurfaces \cite{oneill1983semiriemannian}.

For the metric (\ref{frw}) and the frame (\ref{thesolution}), the gravitational field strength tensor as defined at (\ref{GPReq}) is given by
\bs
\be 
F_\alpha{}^{\mu\nu} = 2\kappa H\lp 1 + \frac{k}{a^2H}\rp h_\alpha{}^{[\mu}u^{\nu]}\,,
\ee
or more explicitly, 
\be
F_{a}{}^{\mu\nu}=2\kappa\lp H^2 + a^{-2}k\rp\,u^{[\mu}\pi_{a}^{\nu]}\quad \text{where} \quad
\pi_{0}^{\mu}=-\chi  x^{i}(\partial_{i})^{\mu}\,,\quad
\pi_{i}^{\mu}=\ell \big(\delta_{i}{}^{j}-k x_{i}x^{j}\big)(\partial_{j})^{\mu}\,.
\ee
\es
These expressions neatly capture the purely `electric' property of the gravitational field in cosmology through the structure $\sim u^{[\mu}\pi^{\nu]}_a$. `Magnetic' components $\sim \pi^{[\mu}_a\pi_b^{\nu]}$ are absent due to isotropy since they would introduce a preferred spatial direction. 
The energy excitation $F_0{}^{\mu\nu}$ describes radial Hubble flow, the spatial vector $\pi_{0}^{\mu}$ encoding the energy transport wrt the canonical time direction.  The momentum excitations $F_i{}^{\mu\nu}$ involve the principal directions $\pi_{i}^{\mu}$ and describe the expansion-induced deformation of the canonical transvections.   

Let us compute the charges (\ref{GPRcharge}) in a spherical region within the radius $r < R$. The unit normal $n$ is then $n_i = a\chi^{-1}x_i/r$, and the induced line element 
on the surface of the sphere is most conveniently written in spherical coordinates, $\sqrt{\gamma}\diff^2 x = a^2 r^2\sin\theta\diff\theta\diff\phi$. We arrive at the charges
\be \label{quasicharge}
q_0 = 4\pi  \kappa\lp aR\rp^3\lp H^2 + \frac{k}{a^2}\rp\,, \quad q_i = 0\,.  
\ee
The energy coincides with the Misner-Sharp mass. This is the physically expected result. However, mathematically it appears to provide a nontrivial consistency check of the canonical Noether construction, despite the earlier observation that $\ie_0=K$. Though computed for the same generator, the charge is obtained from a different superpotential than the Komar-Kodama-Hayward 2-form $\omega$ \cite{Hayward:1994bu} 
\be
\omega^{\mu\nu} =  -2\kappa\mathring{\nabla}^{[\mu}K^{\nu]} = -2\kappa X^{[\mu\nu]\alpha}K_{\alpha}\,.
\ee
To rewrite the $\omega$ in the second form, we used the frame-compatibility of the $\nabla$ and the definition of the X-tensor (\ref{GPRaction}). The $\omega$ is a different contraction than the $F_0$, but apparently the difference turns out to be a total derivative under sufficiently symmetric conditions. 

Notice that so far we have only used the first equation (\ref{GPReq}). In particular, we have not introduced any matter sources nor considered any dynamical equations for the metric. The canonical frame and the physical energy-momentum is determined in G$_\parallel$R solely from the metric geometry.  

To finally report the conserved material currents, we introduce a perfect fluid energy-momentum tensor $T^{\mu\nu}$ with the energy density $\rho = u_\mu u_\nu T^{\mu\nu}$ and the pressure $p = h_{\mu\nu}T^{\mu\nu}/3$. The canonical currents then promptly follow by using the solution (\ref{canonicalframe}), 
\bs
\label{Vcurrents}
\ba
J_0 & = & \chi\lp \rho u + H p x^i\partial_i\rp\,,  \\
J_i & = & \ell k\rho x_i u - \ell Hp\lp \delta^j_i -kx^jx_i\rp\partial_j\,. \label{Ji}
\ea
\es
The energy current $J_0$ contains the energy density $\rho$ and acquires a spatial contribution proportional to $H p x^{i}\partial_{i}$, reflecting the fact that in an
expanding universe pressure transports energy radially through the canonical frame. When $k\neq 0$, the momentum currents $J_i$ contain the momentum density $\rho x_i$, and a pressure-induced shear term proportional to $-Hp$, mirroring the `redshifting' of the transvection generators (\ref{0icommutator}). 
The arbitrary length scale $\ell$ remains in the expression for the momentum currents, but drops out of all integrated charges, leaving the observables unaffected. Whilst the interpretation of the scale $\ell$ in a generic frame field (\ref{regular}) is less transparent, in the canonical frame (\ref{thesolution}), it simply sets the length unit for the spatial triad. 

It is straightforward to verify that the four $J_a$ are all conserved, given the continuity equation $\rho' + 3H(\rho +p)=0$. The latter is implied by the second equation (\ref{GPReq}). The integrated charges according to (\ref{Vcharge}) obtained from the currents (\ref{Vcurrents}) are $Q_0 = \rho V$, $Q_i=0$, where $V=4\pi(aR)^3/3$.  
By the Friedmann constraint $3\kappa (H^2 +k/a^2)= \rho$, which also follows from the second equation (\ref{GPReq}), we see that $Q_a=q_a$, as was expected in the absence of topological subtleties. As anticipated in the discussion of the two charge definitions (\ref{NVcharge}), adopting the covariant (\ref{Vcharge}) was
essential to cancel the lapse dependence from the observables.

\section{Conclusion}
\label{conclu}

In G$_\parallel$R the energy-momentum charges arise as covariantly defined Noether charges, determined solely by the spacetime geometry. 
No Killing symmetries are required. 
Beginning with the disambiguation of the coordinate system and the reference frame, we promoted the latter into a field subject to a boundary-restricted variational principle. 
By requiring the variations of the frame field to be restricted at the boundary, and demanding that the frame itself carries no energy, we arrived at an essentially unique theory, encapsulated in the two equations (\ref{GPReq}).   

Using cosmology as an explicit example, we illustrated how the theory selects a canonical frame. The Kodama vector $K$, which in GR appears as an exceptional construction tied to spherical symmetry, is recovered here as the canonical time-evolution generator $\ie_0$, a component of the frame field that arises as a solution of the G$_\parallel$R equations. The associated energy is obtained as the $\ie_0$-generated Noether charge and in cosmology coincides with the Misner-Sharp mass, while the full set of conserved currents $J_a$ yields a complete and consistent description of energy and momentum flows in an expanding universe. 
  
Black hole spacetimes provide a complementary application, probing qualitatively different aspects of the construction. 
In this setting, the excitation $F_0{}^{\mu\nu}$ correctly captures the gravitational flux even in the absence of sources, so that $q_0 \neq Q_0$ due to the presence of a singularity. Moreover, the black hole partition function is obtained from the action without the need to introduce boundary terms or counterterms, since in this framework the physical partition is fixed by the canonical frame rather than by a choice of renormalisation scheme \cite{BeltranJimenez:2024ufa}.  
With the canonical energy, momenta, and angular momenta well-defined in generic spacetimes, these results motivate applications of G$_\parallel$R beyond the idealised settings considered so far\footnote{It is worth noting that the explicit construction of solutions can be facilitated by geometric methods developed in the context of affine geometry enriched by torsion and non-metricity, properties of the frame-covariant $\nabla$ that also introduce 
an additional layer of interpretation to the theory, see e.g. \cite{BeltranJimenez:2019esp}. This perspective, while natural within the present framework, was deliberately left outside the scope of this paper.}. 
 
A canonical geometric notion of energy appears to be a natural prerequisite for a consistent general-relativistic quantum theory, since the time evolution and physical interpretation of quantum states rely on a Hamiltonian.
While quantum superposition and entanglement are inherently frame-dependent, in non-gravitational physics this dependence can be anchored to a fixed background structure: Galilean time in quantum mechanics and Minkowski spacetime in quantum field theory. In the presence of dynamical geometry, this anchoring is lost. The canonical frame in G$_\parallel$R offers a concrete geometric foundation for defining general-relativistic quantum states and observables.

\acknowledgements{This work was supported by the Estonian Research Council through the grants PRG2608 and TK202. 
Part of this work grew out of a collaboration with D{\'e}bora Aguiar Gomes and Jose Beltr{\'a}n Jim{\'e}nez. }

\bibliography{paracosmo}

\end{document}